\pdfoutput=1

\documentclass[11pt]{article}
\usepackage{booktabs}

\usepackage[preprint]{acl}

\usepackage{times}
\usepackage{latexsym}

\usepackage[T1]{fontenc}

\usepackage[utf8]{inputenc}

\usepackage{microtype}

\usepackage{inconsolata}

\usepackage{graphicx}

\title{Comprehensive Audio Query Handling System with Integrated Expert Models and Contextual Understanding}

\author{Vakada Naveen ,
  Arvind Krishna Sridhar,
  Yinyi Guo ,
  Erik Visser  \\
  Qualcomm Technologies  
  }

\begin{document}
\maketitle
\begin{abstract}
 This paper presents a comprehensive chatbot system designed to handle a wide range of audio-related queries by integrating multiple specialized audio processing models. The proposed system uses an intent classifier, trained on a diverse audio query dataset, to route queries about audio content to expert models such as Automatic Speech Recognition (ASR), Speaker Diarization, Music Identification, and Text-to-Audio generation. A 3.8 B LLM model then takes inputs from an Audio Context Detection (ACD) module extracting audio event information from the audio and post processes text domain outputs from the expert models to compute the final response to the user. We evaluated the system on custom audio tasks and MMAU sound set benchmarks. The custom datasets were motivated by target use cases not covered in industry benchmarks and included ACD-timestamp-QA (Question Answering) as well as ACD-temporal-QA datasets to evaluate timestamp and temporal reasoning questions, respectively. First we determined that a BERT based Intent Classifier outperforms LLM-fewshot intent classifier in routing queries. Experiments further show that our  approach significantly improves accuracy on some custom tasks compared to state-of-the-art Large Audio Language Models and outperforms models in the 7B parameter size range on the sound testset of the MMAU benchmark,  thereby offering an attractive option for on device deployment. 
\end{abstract}

\section{Introduction}
The rapid advancement of LLMs has significantly transformed the capabilities of chatbot systems, particularly in handling text-based queries \cite{kumar2023large}. However, the domain of audio content related queries remains relatively underexplored, with existing chatbots often limited to specific audio tasks \cite{microsoft2015luis,google2010dialogflow,aws2017lex,chu2023qwen,tang2023salmonn}. There is an  increasing demand for intelligent systems capable of processing and understanding audio data in various contexts \cite{zhao2019applications}. Whether it is recognizing music tracks, transcribing spoken language, or identifying speakers in a conversation, the ability to accurately interpret audio inputs is crucial for enhancing user interaction and satisfaction. Traditional chatbots like LUIS \cite{microsoft2015luis}, DialogFlow \cite{google2010dialogflow} and Lex \cite{aws2017lex} , which primarily focus on speech or text, fail to meet these needs, highlighting the necessity for a more comprehensive approach. Even the latest multimodal LLMs are targeted towards specific speech and general audio related tasks \cite{tang2023salmonn,chu2023qwen,gong2023listen,wu2023decoder} fail to answer diverse audio queries.

This paper introduces a novel chatbot system designed to address a broad spectrum of audio-related queries by integrating multiple specialized audio processing models. Our goal is to create a versatile and robust solution that surpasses the limitations of current systems. To this end, we first developed an intent classifier that effectively routes user queries to the appropriate audio expert models. This classifier was trained on a diverse dataset of audio-related questions, ensuring it can handle a wide range of queries with high accuracy. By leveraging advanced models such as Automatic Speech Recognition (ASR) \cite{malik2021automatic}, Speaker Diarization \cite{park2022review} , Music Identification \cite{chaouch2020audio}, and Text to Audio generation \cite{huang2023make}, our system can process and respond to complex audio queries that require a combination of one or more of these expert models. Moreover, our system incorporates language models to provide coherent and contextually relevant responses. These models integrate text outputs from the audio expert models with additional context from the Audio Context Detection (ACD) model \cite{kong2020panns}, which predicts audio events present in an audio along with their timestamps. This integration is particularly beneficial for handling multifaceted queries that require a deep understanding of the audio context.

Experimental results highlight the effectiveness of our approach. A BERT-based intent classifier is shown to outperform a LLM-Fewshot intent classifier in terms of precision, recall, and F1-score. Furthermore, incorporating explicit ACD metadata in JSON format as inputs to the LLM significantly boosts the system’s accuracy. To gauge the system’s ability to address target use cases outlined in the paper, we also propose 2 new temporal task datasets and assess temporal reasoning skills using various methods, including zeroshot and few-shot prompting with both ground truth and predicted ACD audio events Finally, we compare our model with other large audio language models on the sound test of the MMAU benchmark \cite{sakshi2024mmau} , demonstrating its competitive performance despite lower complexity.

To our knowledge, we are the first to propose a comprehensive solution for audio-related queries by integrating specialized audio processing and advanced language models. In addition we introduce a novel audio intent dataset for training a BERT based query intent classifier and assess system performance on  two benchmark datasets for temporal reasoning tasks beyond what is covered in MMAU sound set.

\section{Related Works / Background}

\subsection{Large Audio Language Models}

Recent studies have explored the integration of speech signals into large language models (LLMs), enabling them to directly process and understand general audio inputs. SALMONN \cite{tang2023salmonn} is a speech audio language music open neural network that integrates a pre-trained text-based LLM with speech and audio encoders into a single multimodal model, achieving competitive performance on various speech and audio tasks. Similarly, the Qwen-Audio \cite{chu2023qwen} model scales up pre-training to cover over 30 tasks and various audio types, enhancing interaction capabilities with Qwen-Audio-Chat for multi-turn dialogues. The LTU (Listen, Think, and Understand) model \cite{gong2023listen} introduces a new approach by combining audio perception with reasoning abilities, trained on the OpenAQA-5M dataset to exhibit strong performance on classification and captioning tasks, as well as emerging audio reasoning and comprehension abilities. GAMA, a general purpose Large Audio-Language Model (LALM) \cite{ghosh2024gama} , integrates multiple audio representations and fine-tunes on a large-scale audio-language dataset, augmented with complex reasoning abilities through instruction-tuning with a synthetically generated CompA-R (Instruction-Tuning for
Complex Audio Reasoning) dataset, demonstrating advanced audio understanding and reasoning capabilities.

Another approach to integrating speech signals into LLMs is the use of decoder-only architectures, such as Speech-LLaMA \cite{wu2023decoder}. This model leverages a decoder-only architecture to map compressed acoustic features to the continuous semantic space of the LLM, demonstrating a significant improvement over strong baselines on multilingual speech-to-text translation tasks. Furthermore, LauraGPT \cite{du2023lauragpt} is a unified audio-and-text GPT-based LLM that can process both audio and text inputs and generate outputs in either modalities.These advances in audio-text integration demonstrate the potential of LLMs in handling audio-related tasks.

\subsection{Chatbot Systems}

Our work builds upon the recent advances in large language models (LLMs) and their applications in various domains, particularly in the area of multimodal reasoning and action. Recent studies have shown that multimodal LLMs can be used for various audio and vision tasks, such as audio generation and editing \cite{liang2024wavcraft}, speech recognition \cite{huang2024audiogpt}, and image generation \cite{suris2023vipergpt}. Additionally, the concept of generalist agents, which aim to combine basic skills to solve complex tasks \cite{ge2024openagi}, has been explored in the context of LLMs. Furthermore, the use of LLMs to execute computer tasks guided by natural language \cite{kim2024language} has been demonstrated, and open-source AGI platforms that integrate LLMs with domain-specific expert models to solve complex tasks \cite{deng2024mind2web} have been proposed. However, these works do not specifically address the challenge of handling diverse audio-related queries, which is the focus of our paper.

\subsection{Intent classification}

Intent classification entails identifying the main objective or intent of a particular text. There are several datasets used for this task. Specifically, SNIPS \cite{coucke2018snips} dataset covers various domains like music, weather, and booking services, while ATIS \cite{hemphill1990atis} is focused on the air travel domain. The Banking dataset \cite{casanueva2020efficient} is specific to banking and finance, and the Massive dataset \cite{fitzgerald2022massive} spans multiple languages and domains. The SLURP dataset \cite{bastianelli2020slurp} is a large and diverse multi-domain dataset for end-to-end spoken language understanding, featuring around audio recordings annotated with scenarios and actions. Studies \cite{larson2019evaluation} have shown that transformer based BERT based models give the best performance. Recently LLM based approaches \cite{benayas2024enhancing,parikh2023exploring,loukas2023making} have also been proposed which involves various methods of few shot prompting.

\subsection{Expert Audio Models}

A variety of expert models have been deployed for specialized audio tasks and we narrow the description here to a few models explored and aligned with the use cases targeted in this work. Whisper \cite{radford2023robust} is utilized for Automatic Speech Recognition (ASR), while the ACRCloud API \cite{acrcloud} is applied for music identification and recommendation tasks. Pyannote \cite{bredin2023pyannote} is used for speaker diarization, and the VoiceFilter model \cite{wang2018voicefilter} is implemented for personalized speech separation and removal. Additionally, the Audio Question Answering (AQA)-LLM \cite{sridhar2024enhancing} is deployed for answering general audio event-related questions. The Audio Context Detection(ACD) model is based on CNN10 PANN \cite{kong2020panns,mahfuz2023improving} pretrained for audio tagging. The detected audio events and timestamps are determined after applying thresholds on the frame-level probabilities of the model output.

\section{Proposed Method}

The proposed chatbot is illustrated by Figure 1 and is based on determining the intent contained in audio related text queries to then route them to the appropriate audio models. Its individual modules are described in the following sections.
\begin{figure}[ht]
    \centering
    \includegraphics[width=0.9\linewidth]{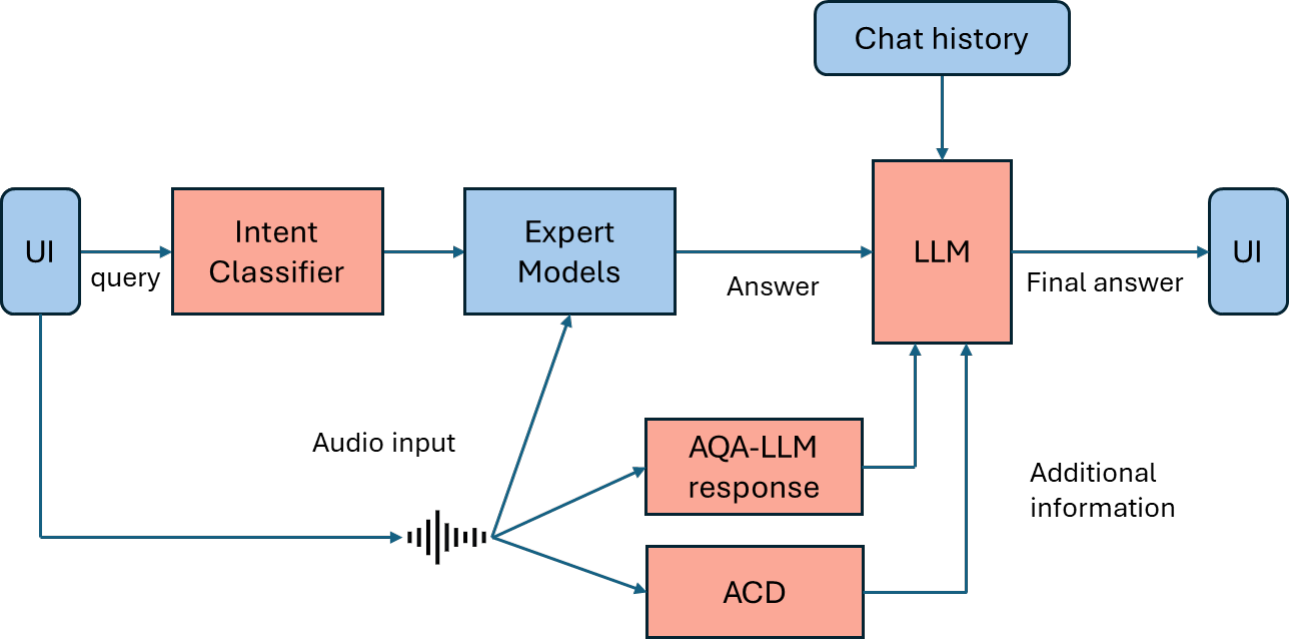}
    \caption{Proposed chatbot system}
    \label{fig:system_block_diagram}
\end{figure}

\subsection{Intent Classifier}

\begin{table}[h!]
\centering
\resizebox{0.5\textwidth}{!}{%
\begin{tabular}{l r r r}
\toprule
\textbf{Class} & \textbf{Training Count} & \textbf{Test Count} & \textbf{Total Count} \\
\midrule

Audio/Text to Audio & 1909 & 478 & 2387 \\
LLM & 1893 & 473 & 2366 \\
Music recommendation & 917 & 230 & 1147 \\
ASR whisper & 778 & 194 & 972 \\
Music identification & 643 & 161 & 804 \\
Speaker ID, Diarization, counting & 602 & 150 & 752 \\
Source separation/removal & 243 & 61 & 304 \\
Unsupported & 2343 & 586 & 2929 \\
\midrule
\textbf{Total} & \textbf{10328} & \textbf{2333} & \textbf{12661} \\
\bottomrule
\end{tabular}%
}
\caption{Proposed Dataset for audio intent classification}
\label{tab:training_test_data}
\end{table}

To build an intent classifier for handling audio queries, we conducted a survey to collect a diverse set of questions related to audio. These questions were then classified to create a robust dataset, which was used to train the intent classifier. 
Our question crowd sourced dataset was acquired through a survey involving 150 participants. This approach was chosen to ensure the data reflects real-life questions people may ask, rather than relying on potentially less authentic open-source datasets. Notably, there are no open-source human generated datasets available for audio intent classifications that cover the queries for the wide range of expert models mentioned in the previous section, making our dataset unique. The dataset is divided into training and test sets, as detailed in Table \ref{tab:training_test_data}, with a total of 12,661 entries, including 10,328 for training and 2,333 for testing. Table \ref{tab:intent_datasets} shows a comparison with other intent classification datasets.

\begin{table}[h!]
\centering
\resizebox{0.38\textwidth}{!}{%
\begin{tabular}{lccc}
\hline
\textbf{Dataset} & \textbf{\# Training} & \textbf{\# Testing} & \textbf{\# Intents} \\
\hline
SNIPS & 13084 & 1400 & 7 \\

ATIS & 4455 & 1373 & 17 \\

Banking & 10003 & 3080 & 77 \\

Massive & 11514 & 3974 & 60 \\

SLURP & 11514 & 2974 & 18 \\
\hline
\textbf{Audio-Intent (Ours)} & \textbf{10382} & \textbf{2333} & \textbf{8} \\
\hline
\end{tabular}%
}
\caption{Intent Classification Datasets }
\label{tab:intent_datasets}
\end{table}

The intent classes include Audio/Text to Audio, LLM, Music recommendation, ASR whisper, Music identification, Speaker ID, Diarization, counting, Source separation/removal, and Unsupported (to classify unsupported tasks). We evaluated the performance of two models: a Bert-based intent classifier and a LLM-Fewshot intent classifier. Even though using LLMs for intent classification can be resource-intensive for relatively simple tasks and may not be suitable for real world applications. we evaluated a fewshot classification approach using the Phi-3.5 model \cite{abdin2024phi} to ensure a comprehensive evaluation.

\begin{table*}[h!]
\centering
\resizebox{0.74\textwidth}{!}{%
\begin{tabular}{lcccccc}
\toprule
\textbf{Class Name} & \multicolumn{3}{c}{\textbf{Bert-based Intent Classifier}} & \multicolumn{3}{c}{\textbf{Phi 3.5 - Fewshot Intent Classifier}} \\
\cmidrule(lr){2-4} \cmidrule(lr){5-7}
 & \textbf{Precision} & \textbf{Recall} & \textbf{F1-score} & \textbf{Precision} & \textbf{Recall} & \textbf{F1-score} \\
\midrule
Music identification & 0.93 & 0.99 & 0.96 & 0.63 & 0.84 & 0.72 \\
Music recommendation & 0.84 & 0.87 & 0.86 & 0.88 & 0.61 & 0.72 \\
ASR whisper & 0.78 & 0.75 & 0.77 & 0.36 & 0.54 & 0.43 \\
Speaker ID, Diarization, counting & 0.78 & 0.89 & 0.83 & 0.72 & 0.68 & 0.70 \\
Audio/Text to Audio & 0.92 & 0.89 & 0.90 & 0.86 & 0.37 & 0.52 \\
Source separation/removal & 0.87 & 0.92 & 0.90 & 0.20 & 0.89 & 0.32 \\
LLM & 0.85 & 0.83 & 0.84 & 0.09 & 0.09 & 0.09 \\
Unsupported & 0.84 & 0.80 & 0.82 & 0.18 & 0.17 & 0.18 \\
\midrule
Overall accuracy & \multicolumn{3}{c}{\textbf{0.85}} & \multicolumn{3}{c}{0.37} \\
\bottomrule
\end{tabular}%
}
\caption{Performance metrics for Bert-based and LLM-Fewshot intent classifiers.}
\label{tab:combined_performance}
\end{table*}

The evaluation metrics include Precision, Recall, and F1-score for each class. Table \ref{tab:combined_performance} presents the detailed performance metrics for both models. The results show that the Bert-based intent classifier gives better results over all these metrics when compared to the fewshot LLM intent classifier. Hence we use the Bert-based intent classifier which is also much smaller than the LLM intent classifier in the proposed system.

\subsection{Expert models}
The trained BERT intent classifier next routes queries to expert audio task models. Our system incorporates several such models.The Automatic Speech Recognition (ASR) model, exemplified by Whisper \cite{radford2023robust}, converts spoken language into text, enabling the system to understand and process verbal queries. The Speaker Diarization model, such as Pyannote \cite{bredin2023pyannote}, identifies and segments different speakers in an audio file, which is particularly useful in multi-speaker environments. For music-related queries, the Music Identification model, like ACRcloud \cite{acrcloud}, recognizes and provides information about music tracks. Additionally, the Text to Audio model generates audio content from text inputs, facilitating the creation of audio responses \cite{huang2023make}.
 The Audio Question Answering Large Language Model (AQA-LLM) \cite{sridhar2024enhancing} is designed to answer questions related to audio events, while the VoiceFilter model \cite{wang2018voicefilter} handles target source separation and removal tasks. Details of the expert model complexities and the deployment can be found in Appendix F and G.

\subsection{Response generation with RAG based LLM}
In the final stage, an LLM combines answers from the expert models  with  chat history and additional inputs from the Audio Context Detection (ACD) model and the AQA-LLM. The chat history consists of the all the conversation happened till the previous turn. We limit the chat history to the last 10 turns since the context length since we have seen degradation of the model performance as the input tokens increases. We chose Phi-3.5 LLM because of its competitive performance on various NLP benchmarks and a 3.8-B parameter model can be deployed on edge devices like mobile phones after further optimizations \cite{abdin2024phi}.

The Audio Context Detection (ACD) model \cite{kong2020panns,mahfuz2023improving} provides meta data  containing detailed event and timestamp information. This model enhances the understanding of the audio environment, allowing the system to generate more accurate and contextually appropriate responses, particularly for queries that require precise timestamp-related information.

To ensure robustness and user satisfaction, our system includes a fallback mechanism. When the expert models are unable to address a query, the AQA-LLM model generates generic answers. This fallback mechanism ensures that the system can handle a wide range of queries, maintaining a high level of user satisfaction even in scenarios where specific expert models may not have the required information. Some examples generated by the proposed chatbot can be found in Appendix E. 


\section{Experimental validation}

\subsection{Custom audio benchmark task datasets}

The two proposed temporal benchmark datasets have two primary applications. Firstly, these datasets are valuable for evaluating Retrieval Augmented Generation (RAG) based approaches, where audio information is represented in text format and text-only LLMs are used to answer audio queries \cite{sakshi2024mmau}. This is facilitated by providing the ground truth ACD metadata in text format along with the QA pairs. Secondly, these datasets are used to directly assess explicit timestamp retrieval  and temporal understanding capabilities required for audio scene understanding in security and monitoring applications for example \cite{sridhar2024enhancing}.

An essential category of questions in our audio chatbot system pertains to timestamp inquiries based on ACD metadata or speaker diarization. To address this, we developed the ACD-timestamp-QA dataset, which comprises 960 entries generated using GPT-4 data augmentation techniques \cite{gong2023listen} to create QA pairs from ground truth audio events data about timestamp information. We used two different representations for the ACD metadata. The first representation is an explicit string format, where the event name, start time, and duration are provided in natural language sentences. The second representation uses JSON entries, which include the audio event name, start time, end time, duration, and the order of the event, with end time, duration, and order explicitly inserted. Our experiments, detailed in the subsequent section, demonstrate that the JSON format with additional explicit information significantly enhances performance compared to the string format.
We also developed the ACD-temporal-QA benchmark to evaluate the temporal audio skills of our approach. Using the GPT-4 model, we generated 1,500 temporal QA pairs from the ground audio events of the Audioset dataset \cite{gemmeke2017audio} as proposed in \cite{gong2023listen}. These questions encompass queries about the chronological order of audio events. We present the results of our experiments in the next section. The ACD-temporal-QA dataset, on the other hand, consists of QA pairs that demand reasoning about the chronological order and temporal nature of audio events, with answers being either “yes” or “no.”

 We utilized the timestamp dataset to derive better representations for ACD metadata and employed the temporal benchmark to evaluate zero-shot, few-shot, and chain-of-thought (CoT) based methods. We present the results of our experiments in section 4.3.1. More details of the proposed custom benchmark tasks can be found in Appendix B and C.

\subsection{MMAU sound set benchmark}
Our datasets differ from the MMAU test set \cite{sakshi2024mmau} as they include ground truth audio events and their timestamps as metadata, which is crucial for assessing the performance of text-only LLMs in temporal and timestamp reasoning. While the MMAU benchmark also features temporal questions, they are multiple-choice and lack the ground truth audio metadata. We use the MMAU benchmark for comparison with other LALMs only. Sample queries for these proposed datasets can be found in Appendix B and C.

\subsection{Results and Discussion}
\begin{table}[h!]
\centering
\resizebox{0.33\textwidth}{!}{%
\begin{tabular}{lcc}
\toprule
\textbf{Model Name} & \textbf{Metadata Type} & \textbf{Accuracy \%} \\
\midrule
AQA-LLM & String format & 74.27 \\
AQA-LLM & JSON format & 80.21 \\
Phi-3.5 & String format & 89.75 \\
Phi-3.5 & JSON format& \textbf{96.35} \\
\bottomrule
\end{tabular}%
}
\caption{Accuracy of different models with various ACD - Metadata types.}
\label{tab:accuracy}
\end{table}

\begin{table}[h!]
\centering
\resizebox{0.38\textwidth}{!}{%
\begin{tabular}{lll}
\toprule
\textbf{Method} & \textbf{Additional Input} & \textbf{Accuracy} (\%) \\
\midrule
Zeroshot & Ground truth & 71.6 \\
Zeroshot + CoT & Ground truth & \textbf{73.66} \\
Fewshot + CoT & Ground truth & 65 \\
Zeroshot & ACD predictions & \textbf{50.34} \\
Zeroshot + CoT & ACD predictions & 48.54 \\
Fewshot + CoT & ACD predictions & 46.12 \\
\bottomrule
\end{tabular}%
}
\caption{System configuration comparisons on custom datasets}
\label{tab:phi_model_evaluation}
\end{table}
\subsubsection{Phi+ACD configuration comparisons}
We first evaluated the performance of different models using various ACD metadata types on the ACD-timestamps-QA dataset. Specifically, we compared the AQA-LLM model and the Phi-3.5 model using both string format and JSON format with extra information. The results, shown in Table \ref{tab:accuracy}, indicate that the JSON format with additional explicit information significantly improves accuracy. Note that we used the ground truth ACD data as the inputs while evaluating these models.

Table \ref{tab:phi_model_evaluation} presents the evaluation results of the Phi model using different prompting methods \cite{kojima2022large} and additional inputs on the ACD-temporal-QA dataset. The highest accuracy of 73.66\% was achieved using the Zeroshot + CoT method with ground truth audio events. CoT method involves prompting the model to provide explanations along with the answers. This indicates that combining the Chain of Thought (CoT) approach with Zeroshot learning significantly improves performance when accurate audio event data is available. In contrast, the Fewshot + CoT method with ground truth audio events resulted in a lower accuracy of 65\%, suggesting that the Fewshot approach may not be as effective in this context. The fewshot method has 2 example QA pairs in the prompt. Detailed prompts for these approaches can be found in Appendix D. 

When using ACD predictions as additional input, all methods showed a notable decrease in accuracy. The Zeroshot method achieved 50.34\%, while the Zeroshot + CoT and Fewshot + CoT methods resulted in even lower accuracies of 48.54\% and 46.12\%, respectively. This decline highlights the challenges of using predicted data, which may introduce errors that negatively impact the model's performance. Overall, the results suggest that assuming the ACD model predictions are reliable, Zeroshot + CoT is the best prompting method for accurate answers and is henceforth referred to as Phi+ACD in the following subsection.



\subsubsection{Comparison with SOTA models}
Table \ref{tab:qwen_comparison} presents the performance comparison between SOTA Large Audio Language Models, and our proposed approach, Phi+ACD, across the two proposed datasets: ACD-temporal-QA and ACD-timestamp-QA. The 7B GAMA model \cite{ghosh2024gama} achieved the highest accuracy on the ACD-temporal-QA dataset with 57.53\%, followed by the 3.8 B Phi+ACD at 50.34\%, and Qwen at 44.87\%. On the ACD-timestamp-QA dataset, Phi+ACD leads with 37.57\% accuracy, while Qwen \cite{chu2023qwen} and GAMA scored 30.66\% and 28.56\%, respectively. These results indicate that the GAMA model performs best on temporal question answering tasks, while Phi+ACD shows better performance on timestamp-related questions.


Table \ref{tab:mmaubenchmark} summarizes the results of various models on the MMAU benchmark sound test split. Our Phi-3.5 + ACD model achieved an accuracy of 50.75\% and hence performs similarly to the 7B param range  Llama-3Instruct + strong cap \cite{sakshi2024mmau} .  However it is outperformed by much larger models such as GPT-4o + strong cap. (57.35\%) and Gemini Pro v1.5 (56.75\%).

Human performance on the test-mini split is significantly higher at 86.31\%, highlighting the gap between current models and human-level understanding. Among the models, Qwen-Audio-Chat (55.25\%) and Qwen-2-Audio-Instruct (54.95\%) also show strong performance, at the expense of larger model sizes and requiring specialized instruction tuning.



\begin{table}[h!]
\centering
\resizebox{0.40\textwidth}{!}{%
\begin{tabular}{lccc}
\hline
\textbf{Model} & \textbf{Size (Billion Parameters)} & \textbf{Dataset} & \textbf{Accuracy} \\
\hline
Phi+ACD & 3.8 & ACD-temporal-QA & 50.34 \\
Qwen & 8.4 & ACD-temporal-QA & 44.87 \\
GAMA & 7 & ACD-temporal-QA & \textbf{57.53} \\
\midrule
Phi+ACD & 3.8 & ACD-timestamp-QA & \textbf{37.57} \\
Qwen & 8.4 & ACD-timestamp-QA & 30.66 \\
GAMA & 7 & ACD-timestamp-QA & 28.56 \\
\hline
\end{tabular}%
}
\caption{Model Performance on Different Datasets}
\label{tab:qwen_comparison}
\end{table}

\begin{table}[h!]
\centering
\resizebox{0.40\textwidth}{!}{%
\begin{tabular}{lll}
\toprule
\textbf{Name} & \textbf{Size} & \textbf{Accuracy} (\%) \\
\midrule
Random Guess & - & 26.72 \\
Most Frequent Choice & - & 27.02 \\
Human (test-mini) & - & 86.31 \\
\midrule
Pengi & 323M & 6.1 \\
Audio Flamingo Chat & 2.2B & 23.42 \\
\midrule
LTU & 7B & 22.52 \\
LTU AS & 7B & 23.35 \\
MusiLingo & 7B & 23.12 \\
MuLLaMa & 7B & 40.84 \\
M2UGen & 7B & 3.6 \\
GAMA & 7B & 41.44 \\
GAMA-IT & 7B & \textbf{43.24} \\
\midrule
Qwen-Audio-Chat & 8.4B & 55.25 \\
Qwen2-Audio & 8.4B & 7.5 \\
Qwen2-Audio-Instruct & 8.4B & 54.95 \\
SALAMONN & 13B & 41 \\
Gemini Pro v1.5 & - & \textbf{56.75} \\
GPT4o + weak cap. & - & 39.33 \\
GPT4o + strong cap. & - & 57.35 \\
Llama-3-Instruct + weak cap. & 8B & 34.23 \\
Llama-3-Ins. + strong cap. & 8B & 50.75 \\
\midrule
\textbf{Phi-3.5 + ACD} (proposed approach) & 3.8B & \textbf{50.75} \\
\bottomrule
\end{tabular}%
}
\caption{Results on MMAU Sound Test Split}
\label{tab:mmaubenchmark}
\end{table}

\vspace{-7mm}
\section{Conclusion}

This paper introduces a comprehensive chatbot system that integrates multiple specialized audio processing models and advanced language models to handle a wide range of audio-related queries. Our approach demonstrates competitive performance on custom and MMAU sound set benchmarks when compared against similar sized (3-7B param) models, showcasing its effectiveness in addressing complex audio queries with a tractable footprint. We intend to conduct further system optimizations in future work with the goal of deploying the models on devices with real time computational constraints.

\bibliography{acl_latex}

\appendix

\section{Intent classification dataset examples}

\begin{figure*}[h]
    \centering
    \includegraphics[width=0.9\linewidth]{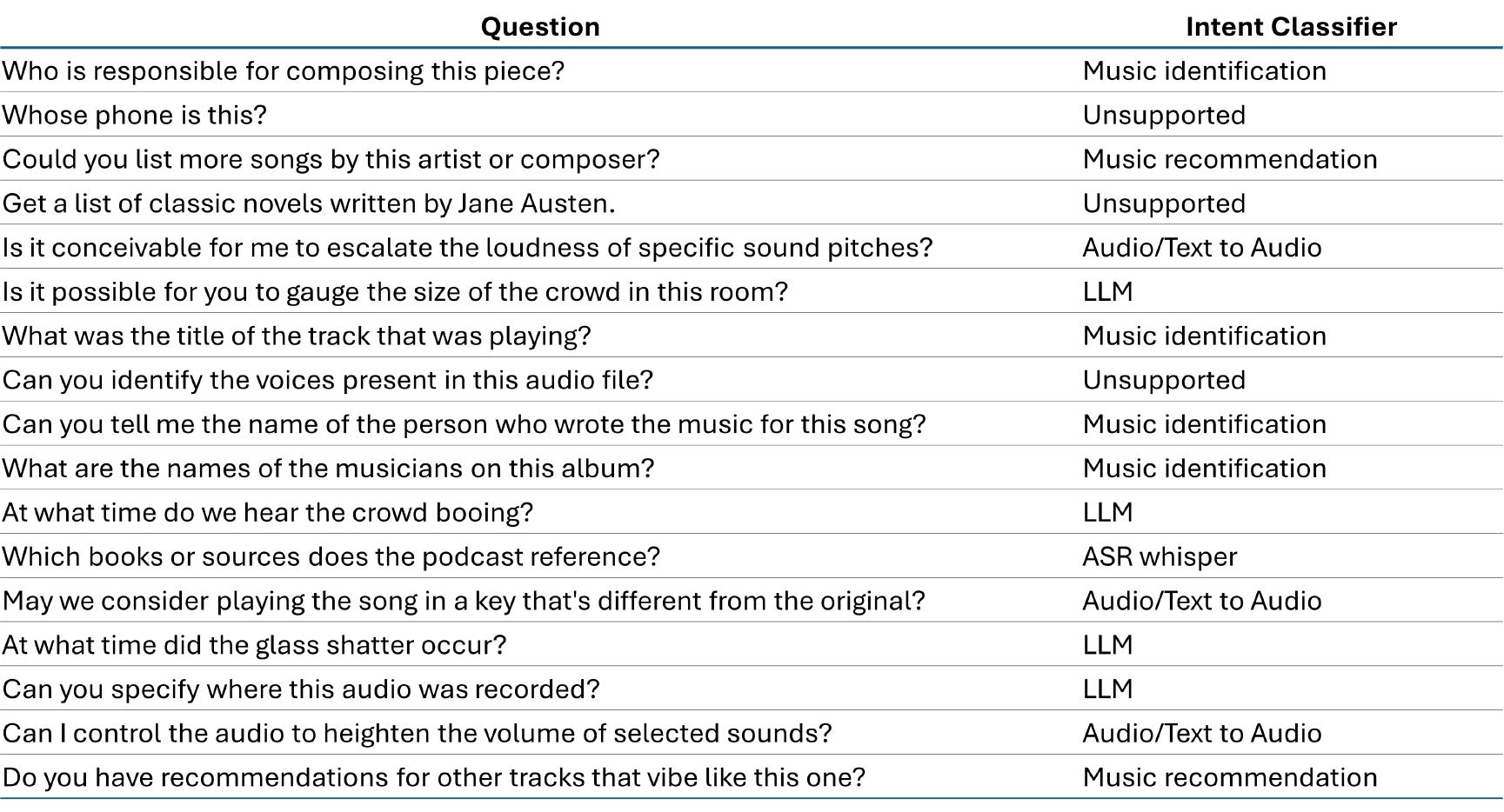}
    \caption{Sample queries from intent classification dataset}
    \label{fig:ic_dataset}
\end{figure*}

Sample queries for each class in the intent classification dataset is shown in Figure \ref{fig:ic_dataset}.

\section{ACD-timestamp-QA dataset examples}
\begin{figure*}[h]
    \centering
    \includegraphics[width=0.5\linewidth]{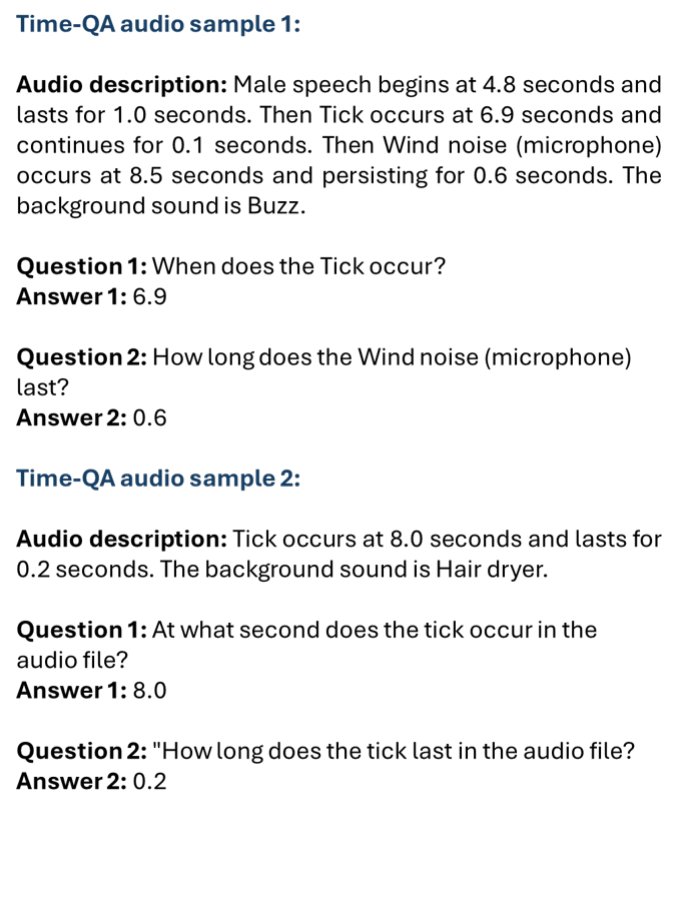}
    \caption{Sample queries from ACD-timestamp-QA dataset}
    \label{fig:timeqa_dataset}
\end{figure*}

Sample queries along with the ground truth ACD metadata for the ACD-timestamp-QA are presented in Figure \ref{fig:timeqa_dataset}.

\section{ACD-temporal-QA dataset examples}
\begin{figure*}[h]
    \centering
    \includegraphics[width=0.5\linewidth]{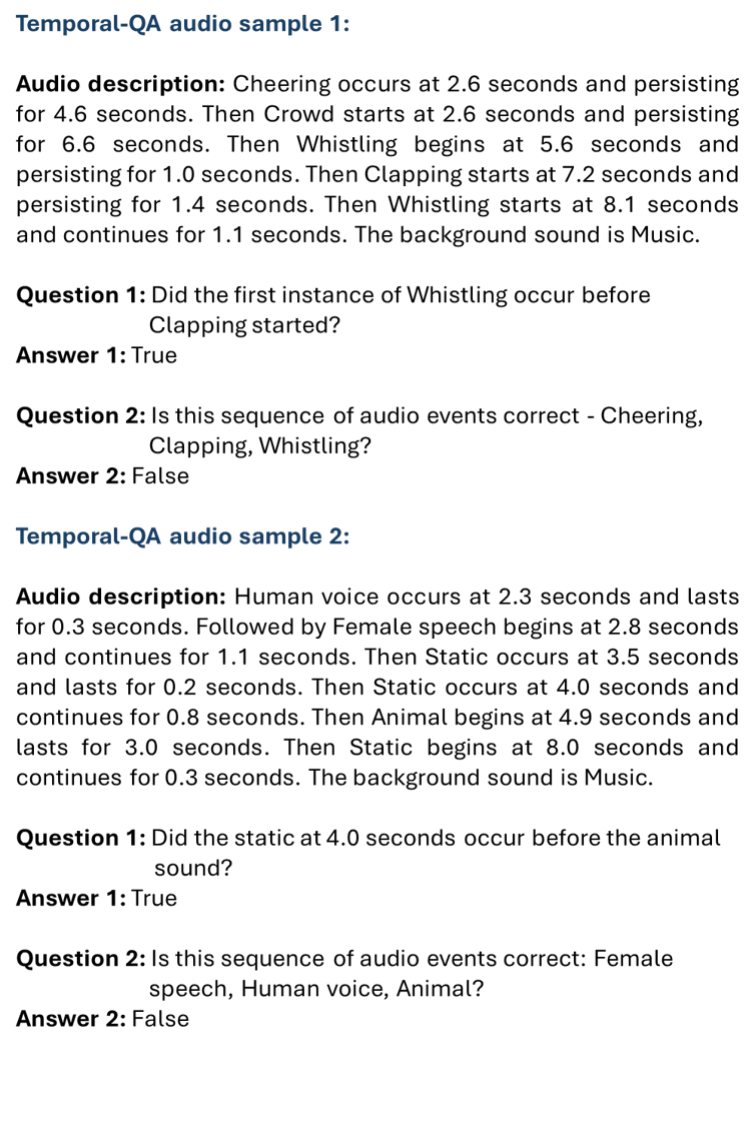}
    \caption{Sample queries from ACD-temporal-QA dataset}
    \label{fig:temporal_dataset}
\end{figure*}

Sample queries along with the ground truth ACD metadata for the ACD-temporal-QA are shown in Figure \ref{fig:temporal_dataset}.

\section{Prompts used for the experiments}
\begin{figure*}[h]
    \centering
    \includegraphics[width=0.7\linewidth]{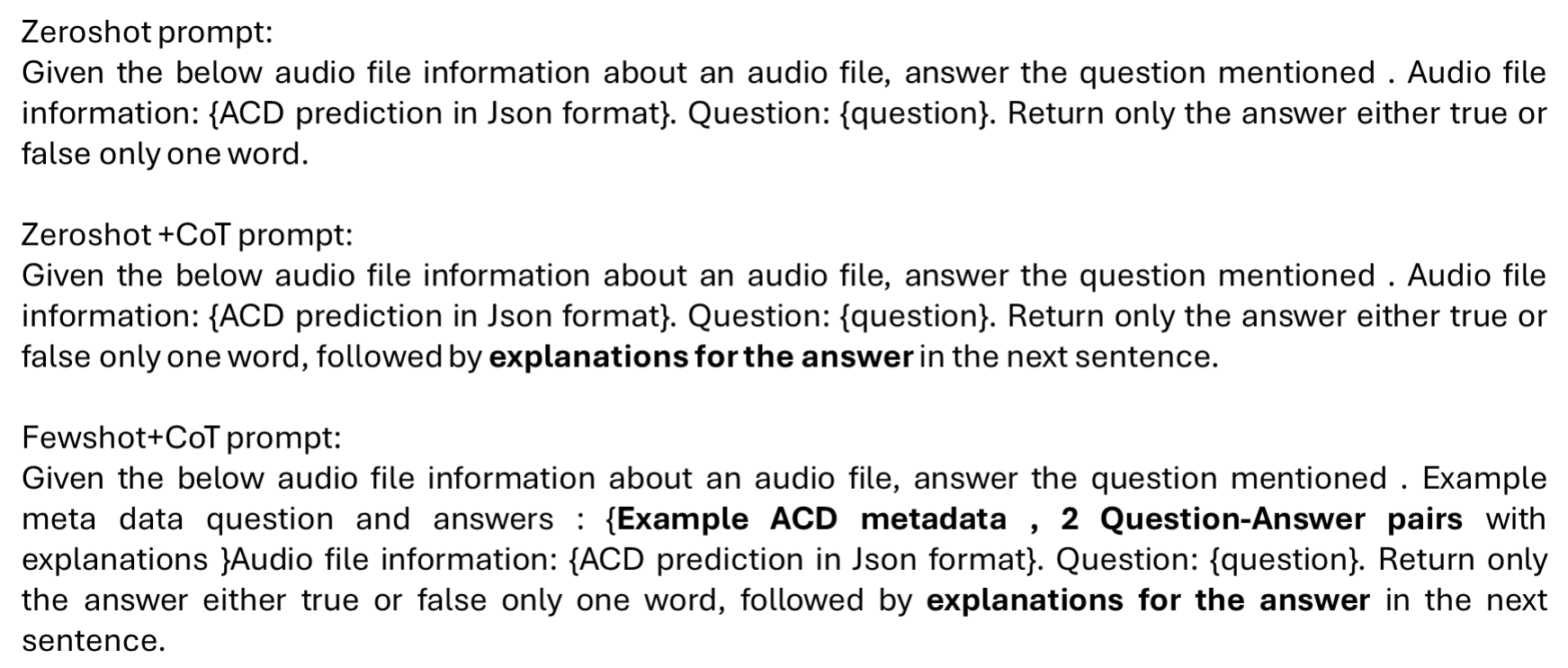}
    \caption{Prompts used for various methods}
    \label{fig:prompt_figure}
\end{figure*}

The prompts used for the experiments in Table \ref{tab:phi_model_evaluation} are presented in Figure \ref{fig:prompt_figure}.

\section{Qualitative examples}
\begin{figure*}[!h]
    \centering
    \includegraphics[width=0.88\linewidth]{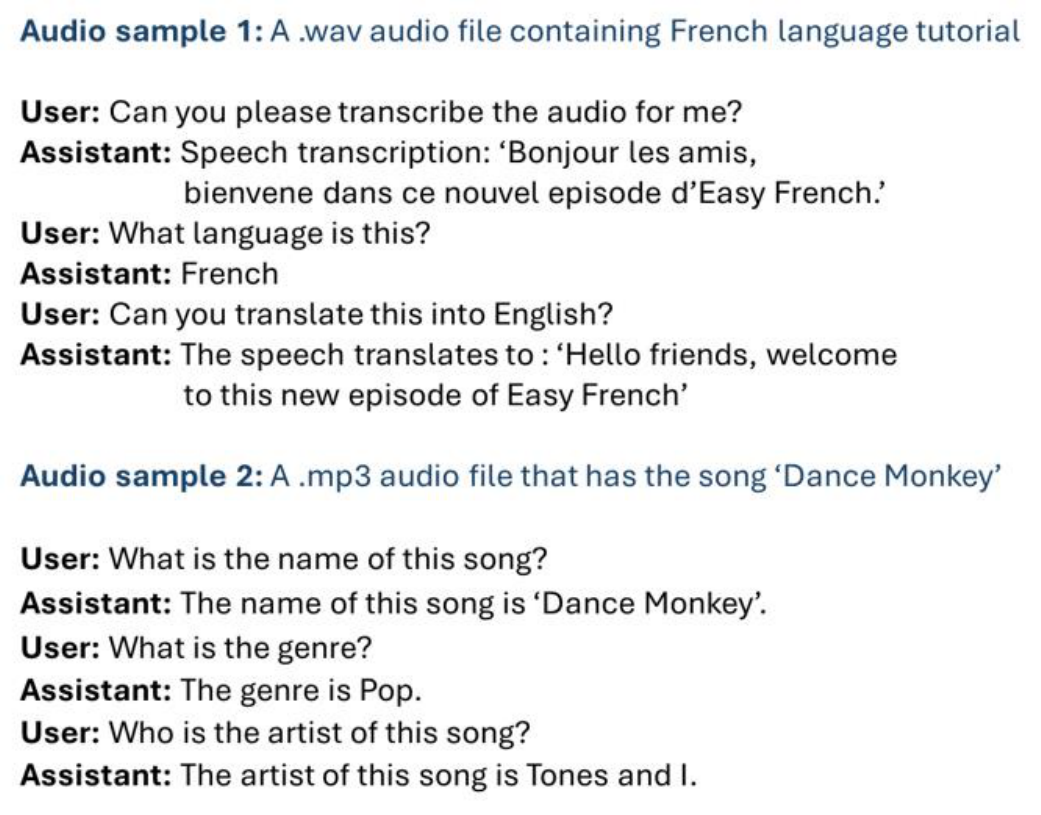}
    \caption{Qualitative examples generated by the proposed chatbot}
    \label{fig:qual_examples}
\end{figure*}

Some of the outputs generated by the proposed chatbot system is shown in Figure \ref{fig:qual_examples}.

\section{On-device and Cloud deployment}
In future , based on resource availability, we propose to deploy the Whisper, ACD, Phi and VoiceFilter models on-device and all other remaining models in the cloud. Our approach of using various expert models gives us the flexibility to adopt such hybrid deployment framework.

\section{Model complexities}

We have customized the expert models with our in-house training data and architectures, hence the model complexity may vary from the open-source implementations. The approximate number of parameters for the expert models we used is shown in Table \ref{tab:model_sizes}
\begin{table}[h!]
\centering
\begin{tabular}{lc}
\hline
\textbf{Model} & \textbf{Size} \\
\hline
ACD & 5.5M \\
ACR-cloud & - \\
Pyannote & 31M \\
VoiceFilter & 6.8M \\
AQA-LLM & 7B \\
Whisper &  39M \\
\hline
\end{tabular}
\caption{Model Sizes}
\label{tab:model_sizes}
\end{table}

\end{document}